\documentclass[usegraphicx,usenatbib,useAMS]{mn2e}
\newcommand\lsim{\mathrel{\rlap{\lower4pt\hbox{\hskip1pt$\sim$}}
        \raise1pt\hbox{$<$}}}
\newcommand\gsim{\mathrel{\rlap{\lower4pt\hbox{\hskip1pt$\sim$}}
        \raise1pt\hbox{$>$}}}

\usepackage{amsmath}
\usepackage{array}

\usepackage{graphicx}
\usepackage{fixltx2e}

\title[${\it H_2}$ self--shielding in 3D Simulations]
{Photodissociation of ${\bf H_2}$ in Protogalaxies:  Modeling Self--Shielding in 3D Simulations}

\author[J. Wolcott-Green et al.]
{J. Wolcott-Green$^{1,3}$, Z. Haiman$^{2}$ and G. L. Bryan$^{2}$\thanks{E-mail: jemma@astro.columbia.edu; zoltan@astro.columbia.edu; gbryan@astro.columbia.edu}\\ 
$^{1}$Columbia Astrophysics Laboratory, 550 West 120th Street, MC 5247, New York, NY 10027, USA\\
$^{2}$Department of Astronomy, Columbia University, 550 West 120th Street, MC 5246, New York, NY 10027, USA\\ 
$^{3}$Barnard College, Columbia University, 3009 Broadway, New York, NY 10027, USA}

\begin{document}

\date{}

\pubyear{2011}

\maketitle

\begin{abstract}
  The ability of primordial gas to cool in proto--galactic haloes
  exposed to Lyman-Werner (LW) radiation is critically dependent on
  the self-shielding of ${\rm H_2}$.  We perform radiative transfer
  calculations of LW line photons, post-processing outputs from
  three-dimensional adaptive mesh refinement (AMR) simulations of
  haloes with $T_{\rm vir} \gsim 10^4$ K at $z \sim 10$. We calculate
  the optically thick photodissociation rate numerically, including
  the effects of density, temperature, and velocity gradients in the
  gas, as well as line overlap and shielding of ${\rm H_2}$ by HI,
  over a large number of sight-lines.  In low--density regions ($n
  \lsim 10^4~{\rm cm^{-3}}$) the dissociation rates exceed those
  obtained using most previous approximations by more than an order of
  magnitude; the correction is smaller at higher densities.  We trace
  the origin of the deviations primarily to inaccuracies of (i) the
  most common fitting formula \citep{DB96} for the suppression of the
  dissociation rate and (ii) estimates for the effective shielding
  column density from local properties of the gas. The combined
  effects of gas temperature and velocity gradients are comparatively
  less important, typically altering the spherically averaged rate
  only by a factor of $\lsim$ two.  We present a simple modification
  to the DB96 fitting formula for the optically thick rate which
  improves agreement with our numerical results to within $\sim 15$
  per cent, and can be adopted in future simulations.  We find that
  estimates for the effective shielding column can be improved by
  using the local Sobolev length.  Our correction to the ${\rm H_2}$
  self-shielding reduces the critical LW flux to suppress ${\rm H_2}$
  cooling in $T_{\rm vir} \gsim 10^4$ K haloes by an order of
  magnitude; this increases the number of such haloes in which
  supermassive ($M \sim 10^5 {\rm \ M_\odot}$) black holes may have
  formed.
\end{abstract}


\section{Introduction}

It has long been known that molecular hydrogen, the most efficient
coolant in metal--free gas at temperatures below $10^4$K, played a
key role in formation of first--generation, ``Population III,'' stars
(see \citealt{AH00} for a review). As soon as these first stars began
to shine, however, they also began to destroy ${\rm H_2}$ via
dissociating (LW) photons in the range 11-13.6 eV, to
which the universe is largely transparent even at early times, $z \sim
20-30$. The nature and extent of this photodissociation feedback has
important consequences for subsequent star-formation, reionization,
and the formation of massive black holes at early times.

In regions where large ${\rm H_2}$ column densities build up ($N_{\rm
  H_2} \gsim 10^{14}~{\rm cm^{-2}}$), photodissociation is suppressed
as the LW bands become optically thick; the cooling
properties of UV--irradiated primordial gas thus depend largely on its
ability to ``self--shield.'' Unfortunately, the problem of modeling
self--shielding exactly in existing studies remains intractable; in
particular, approximate treatments are necessitated by two main
challenges, which are the focus of this paper. First, the
computational expense for three--dimensional simulations of finding
the exact self--shielding column density in a large number of
directions is prohibitive. As a result, studies have typically either
adopted the optically--thin dissociation rate throughout
\citep{MBA01,MBA03,MBH06,MBH09,WA07,WA08a,WA08b,Greif+10}, relied upon
estimating $N_{\rm H_2}$ from local properties of the gas
\citep{BL03,JGB08,GTK09,SBH10,Johnson+11}, or sacrificed angular
resolution, finding the exact column density in a small number of
directions to estimate the dissociation rate
\citep{Yosh+03,Yosh+07,GML07a,GML07b}. Alternatively, some have
employed a local method, which allows for contributions to shielding
only from gas within a single smooth-particle-hydrodynamics (SPH)
smoothing length \citep{GSJ06} or within a width defined by the size
of the underlying simulation grid -- a method also investigated by
\citet{GML07a,GML07b}.  Recently, an algorithm for finding the
projected column density distribution as seen by each SPH particle,
using a Healpix tessellation with 48 equal-area pixels, has been
implemented by \citet{GC11}.  In one-zone models, $N_{\rm H_2}$ must
be specified from only ``local'' properties of the gas, by definition
\citep{ON99,O01,OSH08,SBM10,WGH11}.

Even once an estimate of the self--shielding column is obtained
however, finding the exact photodissociation rate represents a large
computational expense, requiring high numerical resolution in order to
explicitly account for processing of the incident LW radiation as a
function of frequency. Furthermore, even when only the (ortho and
para) ground states of the molecule are populated, there are already a
total of 76 LW transitions that contribute to the total optical depth
in the relevant frequency range (photon energies below 13.6eV). Most
often, studies circumvent this difficulty by adopting analytic
expressions provided by \citet[][hereafter DB96]{DB96} to model
self-shielding, with a few exceptions among semi-analytic models
\citep{HAR00,CFA00,GB01,GB03} and one-dimensional simulations
\citep{Ricotti+01,HI05,HI06}, which include more detailed
calculations.

The two primary goals of this study are (i) to quantify the accuracy
of previous models for self--shielding, and (ii) to provide an
improved analytic fit for the suppression of the photodissociation
rate by shielding, which can be used in future simulations. To
accomplish this, we post--process the outputs from a suite of
simulations performed by \citet[][hereafter SBH10]{SBH10}, who studied
the effects of UV--irradiation on protogalactic haloes with virial
temperatures $T_{\rm vir} \gsim 10^4$K. Within these haloes, we
calculate the exact optically--thick dissociation rate at a large
number of points, in a large number of directions, with a detailed
treatment of radiative transfer, explicitly including the effects of
density, temperature, and velocity gradients in the gas, which are most
often neglected in existing models.

The rest of this paper is organized as follows.  In
\S~\ref{sec:Method}, we describe the SBH10 simulations and our methods
to compute photodissociation rates with shielding.  In this section,
we also recapitulate several of the most common existing methods for
modeling self--shielding.  In \S~\ref{sec:Results}, we present our
main results -- the numerically computed suppression of the
photodissociation rates.  We compare these results to those obtained
with each of the previously used methods, and we also elucidate the
effects of gas temperature and velocity gradients, as well as the
accuracy of the analytic expressions from DB96.  In
\S~\ref{sec:Implications}, we discuss the implications of our results
and the associated uncertainties, and we offer our conclusions in
\S~\ref{sec:Conclusions}.

\section{Numerical Method}
\label{sec:Method}

\subsection{Simulations}

We utilize outputs from a suite of simulations performed by SBH10 with
the Eulerian adaptive mesh refinement + N-body code {\sc enzo}
\citep{Bryan99,NB99,OShea+04}. For the sake of brevity, we limit the
discussion here to the most pertinent features of their numerical
method, and refer the reader to the original study for further
details.

The simulations were performed within a comoving box 1 $h^{-1}$ Mpc on
a side and assuming a $\Lambda$CDM cosmological model with standard
concordance parameters: $\Omega_{\rm DM} = 0.233,~\Omega_b =
0.0462,~\Omega_\Lambda = 0.721,~\sigma_8 = 0.817,~n_s = 0.96,~{\rm
and}~ h = 0.701$. A preliminary run was initialized with a root grid
of $128^3$ and no nested grids. This was performed (with radiative
cooling turned off) in order to identify haloes with virial masses of a
few $\times 10^7 {\rm M}\odot$ at $z \sim 10$. Three of these halo
were then re--simulated at high resolution with new initial conditions
-- three nested grids with an effective innermost resolution of
$1024^3$ -- and with levels of refinement added adaptively. Refinement
was increased when the baryon or dark matter mass exceeded thresholds
of 68 and 683 ${\rm M_\odot}$ respectively, and in order to maintain
sufficient resolution of the local Jeans length -- at least four grid
cells -- to prevent artificial fragmentation. Throughout, the dark
matter (DM) gravity was smoothed on a scale of 0.954 $h^{-1}$
(comoving) parsec.
Each simulation was stopped after reaching a refinement level of 18 --
a resolution of 0.0298 $h^{-1}$ parsec (comoving) or 800 AU (physical).

The non-equilibrium chemistry for a gas of primordial composition was
followed with a chemical network comprising 28 gas--phase reactions,
including ${\rm H_2}$ photodissociation. Radiative cooling by ${\rm
H_2}$ was modeled with the function provided by \citet{GP98}. Each
halo was run with varied LW backgrounds in a range of intensities
$J_{21} = 1 - 10^5$, where the standard normalization is implied here
and throughout, $J_{\rm LW} = J_{21} \times 10^{-21}{\rm ~erg~s^{-1}~
cm^{-2}~sr^{-1}~Hz^{-1}}$, and $J_{\rm LW}$ is the flux intensity at
the average LW frequency (h$\nu = 12.4$ eV). The photo-dissociating
source was modeled as a blackbody with a temperature of ${10^4}$ or
$10^5$K (referred to as T4 and T5 respectively).  Since in the T4
case, the photodissociation of ${\rm H^-}$, rather than direct ${\rm
H_2}$ dissociation, controls the ${\rm H_2}$ chemistry, we restrict
our analysis here to the T5 runs. We defer further details of the
adopted self--shielding model to \S~\ref{sec:approximations}.

\subsection{Numerical self--shielding calculations in 3D}
\label{sec:3Dshielding}

\begin{figure}
  \includegraphics[clip=true,trim=0in 0in 0in 0.4in,
    height=2.8in,width=3.2in]{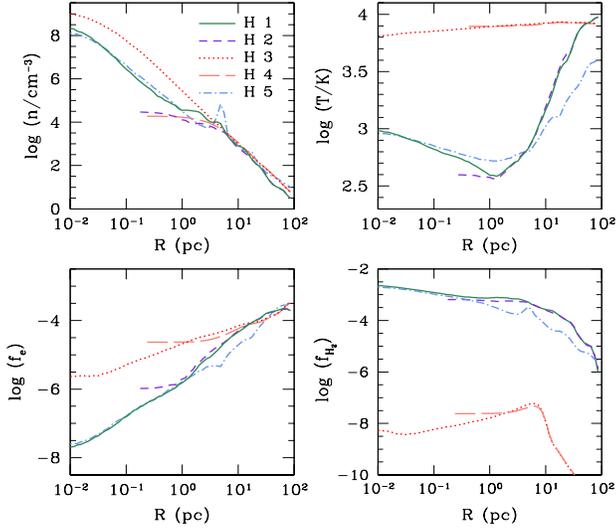}
  \caption{Spherically averaged profiles of temperature, 
    ${\rm H_2}$ fraction, electron fraction, and particle density 
    in the selected simulation outputs. The radius, R, is measured 
    from the densest point in the halo.}
  \label{fig:profiles}
\end{figure}
               
From the simulation outputs, we selected five snapshots spanning a
range in redshift $z \sim 8 - 12$, in which the gas is optically thick
to LW photons. Three of these -- outputs 1, 2, and 5 -- are designated
``cold,'' as they were subjected to a modest dissociating flux, and
thus have experienced significant ${\rm H_2}$--cooling.  In outputs 3
and 4 the gas remains hot, with temperatures $\gsim 7000$ K, having
been exposed to a very strong LW flux that kept the gas ${\rm
H_2}$--poor. The physical properties of the five halo snapshots are
summarized in Table \ref{tbl:HaloProperties}, and the radial profiles
of the density, temperature, ${\rm H_2}$ fraction, and the electron
fraction are shown for each snapshot in Figure~\ref{fig:profiles}.  We
selected $\sim 100$ points from the cold halo outputs at which to
calculate the three--dimensional ${\rm H_2}$--dissociation rate. Fewer
points ($\sim 30$) from the ``hot halo'' outputs were analyzed, as
these are considerably more homogeneous with respect to temperature,
density, and chemical composition (and are nearly optically thin even
in the most dense regions). The selected points span a range in radii
from $0.1 - 10$ pc (physical) where the radius is defined, here and
throughout, as the distance from the densest point in the halo. The
number density and temperature at the selected points vary between
$10-10^{6}~{\rm cm^{-3}}$ and $300-10^4$K, respectively.

\begin{table}
\begin{center}
  \caption{Physical properties of haloes selected from the SBH10 suite
of simulations. Throughout, the outputs will be referenced by number;
1-4 correspond to ``Halo A'' in SBH10, and \#5 to ``Halo C.''  The
redshift ($z$) of each output is given, along with the virial mass at
the collapse redshift (${\rm m _{vir,col}}$; as in SBH10, the virial
mass is defined as the total mass in baryons and dark matter within a
spherically averaged overdensity of 200 with respect to the critical
density). The temperature and total particle densities, $T_0$ and
$n_0$, respectively, are specified at the densest point in the
halo. The intensity of incident Lyman--Werner radiation, $J_{\rm LW}$,
in each run is parametrized in the standard manner: $J_{\rm LW} =
J_{21} \times 10^{-21}~{\rm erg~s^{-1}~cm^{-2}~Hz^{-1}~sr^{-1}}$.}
 
\label{tbl:HaloProperties}
\begin{tabular*}{0.475\textwidth}{@{\extracolsep{\fill}} l l l l l l }
  \hline\hline
  output & $J_{21}$ & $z$ & ${\rm m_{vir,col} (M_\odot)}$ & $T_0 ({\rm K})$ 
& $n_0 ({\rm cm^{-3}}$) \\
  \hline
  1  & $10^2$ & 12.274 & 2.45 $\times 10^7$ & $9.2 \times 10^2$ & $3.1 \times 10^8$ \\
  2  & $10^2$ & 12.285 & 2.45 $\times 10^7$ & $4.0 \times 10^2$ & $3 \times 10^4$  \\
  3  & $10^5$ & 9.93   & 5.49 $\times 10^7$ & $6.3 \times 10^3$  & $1.4 \times 10^9$\\    
  4  & $10^5$ & 9.94   & 5.49 $\times 10^7$ & $7.8 \times 10^3$ & $1.9 \times 10^4$\\ 
  5  & $10^2$ & 8.3    & 7.89 $\times 10^7$ & $9.3 \times 10^2$ & $1.6 \times 10^8$ \\     
  \hline\\
\end{tabular*}
\end{center}
\end{table}

\subsubsection{Rate of photodissociation} 

${\rm H_2}$ is photodissociated primarily via the two-step Solomon
process \citep[Solomon 1965; see also][]{FSD66,SW67}, in which
molecules are excited from the electronic ground state, ${\rm X}^1
\Sigma^+_g$, to the ${\rm B}^1 \Sigma^+_u$ or ${\rm C}^1 \Pi_u$
state\footnote{ ${\rm C}^1 \Pi_u$ is split into $\Pi^+_u$ and
$\Pi^-_u$ states owing to $\Lambda$-doubling, the two-fold degeneracy
of each rotational level ($J$).}, the Lyman and Werner bands
respectively. Subsequent decays lead to the vibrational continuum of
the ground state, rather than to a bound state $\sim 15$ per cent of
the time, thus dissociating the nuclei.

The ``pumping rate'' from a given rovibrational state ($v,J$) to 
the excited electronic state with ($v',J'$) is:
\begin{equation}
\zeta_ {\it{v,J,v',J'}} = \int_{\it{\nu_{th}}}^{\infty}4\pi
\sigma_{\nu}\frac{J_\nu}{h_{\rm P}\nu}{\rm d}\nu,
\end{equation}
where $\sigma_{\nu}$ is the frequency dependent cross-section and
$h_{\rm P}$ is Planck's constant. The frequency threshold, $\nu_{th}$,
corresponds to the lowest energy photons capable of efficiently
dissociating ${\rm H_2}$, with $h\nu \approx 11.1$ eV.  We do not
include flux at $h\nu \geq 13.6~{\rm eV}$, as photons with energies
above the Lyman limit are assumed to have already been absorbed by the
neutral HI in the intergalactic medium (IGM) outside the halo (and
were not included in SBH10).

The dissociation rate from the initial ($v,J$) is then obtained
from the product of the pumping rate and the fraction of decays
leading to the vibrational continuum from $(v',J')$, with a sum 
taken over all possible upper states:
\begin{equation}
k_{{\rm diss},v,J} = \sum_{\it v',J'} \zeta_{\it v,J,v',J'}
{\it f}_ {{\rm diss},v',J'}. 
\end{equation}
The dissociation probabilities here, $f_{{\rm diss},v',J'}$, are
obtained from \citet{ARD00}. The sum over rates from all lower levels,
weighted by the fraction of molecules initially in each, $f_{v,J}$,
then gives the total rate:
\begin{equation}
k_{\rm diss} = \sum_{\it v,J} k_{{\rm diss},v,J} {\it f}_ {v,J},
\end{equation}
where the $f_{v,J}$ are given according to a Boltzmann distribution,
unless otherwise specified.

\subsubsection{Radiative transfer in the haloes}
In the three--dimensional calculations of $k_{\rm diss}$, the first step
is to generate a set of rays\footnote{We use the analysis toolkit {\sc yt} 
\citep{YT10} to interface with the raw simulation data; see 
\S~\ref{sec:ConvergenceTest} for the required angular resolution.} 
emanating from each point where the dissociation rate is to be found.
These sample evenly in the azimuthal angle and in the cosine of the polar 
angle, tiling a sphere of radius $\sim 100$ pc. Along each ray the 
properties of the gas are sampled at intervals determined by the size 
of the underlying grids; the distance between sample points is typically 
0.01 pc in regions with the highest levels of refinement, and 
increases to $\sim 5$ pc toward the outskirts (areas with lowest 
resolution) of the halo.

In our fiducial calculations, the spectrum of the incident radiation
is initialized to be flat in the range 11.1-13.6 eV, with the implicit
assumption that processing of the LW background in the IGM is
negligible. Note that, in general, the cosmological background ${\it
will}$ be modified, though primarily by HI absorption, and not by
${\rm H_2}$ itself (\citealt{HAR00}, hereafter HAR00, see also
\citealt{Ricotti+01}) and the impact of the resulting ``sawtooth''
modulation is considered in \S~\ref{sec:HI}. Tracing a ray from the
outside in, the gas is treated as a series of thin slabs, each with
uniform density, temperature, bulk velocity, and chemical composition
defined at the sample point; the column density is specified by
$n_{\rm H_2} \times \Delta s$, where $\Delta s$ is the width of the
slab. The frequency-dependent optical depth, $\tau_\nu$, of the slab
is then obtained by summing over contributions from all included LW
transitions, each of which is modeled by a Voigt profile. The
rest--frame frequencies are Doppler shifted according to the slab's
line--of--sight velocity relative to the point where the dissociation
rate is to be calculated. The numerical wavelength resolution ($\Delta
\lambda = 2 \times 10^{-4}$\AA~ at the lowest temperatures) is set
adaptively in order to always resolve the thermal line width and is
sufficient to explicitly account for overlap of the Lorentz wings. The
necessary molecular data for these calculations are provided by
\citet{ARLa93,ARLb93}.

We include transitions from the 29 bound rotational levels within the
ground electronic and vibrational ($v=0$) states to excited states
with $v' \leq 37, J \leq 10$; in total, this amounts to 1492 possible
transitions out of $v =0$. We do not include absorption from higher
vibrational levels, which are populated only at particle densities
much larger than we consider here, $n \gsim 10^8~{\rm cm^{-3}}$
(\citealt[][hereafter LPF99]{LPF99}, \citealt{FH07}). In our fiducial
calculations, the rotational levels of $v = 0$ are populated according
to a Boltzmann distribution defined by the temperature of the slab,
irrespective of the local density.  Strictly speaking, populations in
rotational states within $v = 0$ do not thermalize until (temperature
dependent) critical densities are reached, usually taken to be $n_{\rm
crit} \simeq 10^4~{\rm cm^{-3}}$ \citep[but see Table 1 in][]{FH07},
at which depopulation of excited states is dominated by collisional
de--excitation.\footnote{Note that the critical density is different
for each species that perturbs the molecule; here and throughout, we
refer to $n_{\rm crit}$ for collisions with atomic hydrogen only.}
To further address this issue, we perform an additional set of
calculations in which all molecules are assumed to be in the ground
states of para (ortho) hydrogen, $v = 0, J=0~(1)$. In this case, all
76 possible transitions are included; the results are discussed in
\S~\ref{sec:LevelPopulations}.

Finally, having found $\tau_\nu$ along a sightline, the
photodissociation rate is calculated at the point of interest. This
procedure is repeated for each of the sight lines and the final rate
is obtained from a simple average over all directions. Note, however,
that the spherical average is only meaningful in the case of an
isotropic UV background. This would be relevant when the Olber's
integral for the local flux is dominated by a large number of distant
sources.  For haloes with unusually bright and close neighbors, the
flux can be dominated by a single (or a few) of the nearby sources.
In \S~\ref{sec:anisotropy} we discuss a radiation field with a
preferential direction that would be relevant in this case, and
interpret our results in this context.

One remaining caveat is the possibility of ${\rm H_2}$ shielding by
HI.  In general, Lyman--series absorption within the haloes can
suppress the ${\rm H_2}$ dissociation rate by a large factor, though
this requires high optical depth in the wings of the HI lines, and
thus $N{\rm_{HI} \gsim 10^{22}~cm^{-2}}$ \citep{WGH11}. Sufficiently
large neutral column densities are indeed present in the the outputs
we analyze, though only at small radii ($\lsim$ a few pc), and the 
resulting modification of $k_{\rm diss}$ in these regions is discussed 
in \S~\ref{sec:HI}.

\subsection{Approximate treatments of self-shielding}
\label{sec:approximations} 

Approximations for self-shielding in existing simulations have been
necessitated by the two challenges mentioned above: first, the
computational expense for simulations of finding the ${\rm H_2}$
column density in a large number of directions is prohibitive. An
estimate for this inherently non--local quantity therefore typically
must be obtained from purely local information. In
\S~\ref{sec:Napprox}, we give a detailed account of several ways in
which this is commonly achieved. Second, calculating the exact
suppression of the optically--thin rate, with full radiative transfer
in each LW line, is expensive, requiring high numerical wavelength
resolution, as well as the inclusion of non-local effects from the
temperature and velocity structure in the gas. Therefore, studies
often rely on an analytic expression for the optically--thick rate
provided by DB96. We briefly describe this method in
\S~\ref{sec:DB96}, including the assumptions and limitations in
applicability of the analytic fitting formula.

\subsubsection{The self--shielding column density}
\label{sec:Napprox}

In order to estimate the column density, several common methods
make use of local properties of the gas to define a characteristic 
length scale, $L_{\rm char}$; the column density is then obtained 
from:  
\begin{equation}
N_{\rm H_2} = n_{\rm H_2} L_{\rm char},
\end{equation}
with the implicit assumption that the ${\rm H_2}$ number density,
$n_{\rm H_2}$, is constant at the local value over the length $L_{\rm
char}$. Several methods have been used to define a characteristic
length scale: \\

\noindent{\it The Jeans Length} \\ A common approach in both
simulations and one-zone models is to assume the total mass in the
optically--thick region is of order the Jeans mass, with $L_{\rm
char}$ then defined using the local Jeans length. In regions where
particle densities exceed $n{\rm_{tot} \gsim 10^5~cm^{-3}}$, SBH10
have shown that this provides a very accurate estimate of $N_{\rm
H_2}$; however, in lower density regions, it typically overestimates
the integrated $N_{\rm H_2}$ by up to an order of magnitude (see their
Figure 9).  This method also does not account for temperature or
velocity gradients in the gas. \\

\noindent{\it The Sobolev Length} \\
Large velocity gradients will cause the LW resonances, as seen from
a given point in the cloud, to appear shifted from their rest-frame 
wavelength, making the systematic depletion of dissociating photons 
less efficient than if the gas were static. If the fluid motions
are largely disordered and/or supersonic, the exact column density 
along a line of sight can then exceed the effective self-shielding 
column density, $N_{\rm eff}$, by a large factor.

In general, $N_{\rm eff}$ depends on the detailed velocity structure
in the gas; however, it may be estimated by taking a characteristic
length equal to the Sobolev length, $L_{\rm Sob}$ \citep{Sob}.  This
defines a distance over which the mean (macroscopic) fluid velocity
changes by a factor of the (microscopic) thermal velocity of the
molecules, $v_{th}$,
\begin{equation}
L_{\rm Sob} \equiv \frac{v_{th}}{|{\rm d}v/{\rm ds}|}
\end{equation}
with the assumption that ${\rm d}v/{\rm ds}$ is constant. 
Then for a given sightline, $L_{\rm Sob}$ is the distance at which 
the absorption profiles will be shifted by one local line width.
In adopting the Sobolev length for $L_{\rm char}$, it is implicitly
assumed that Doppler shifts in the LW resonances of molecules closer
than $L_{\rm Sob}$ can be ignored, and that all molecules beyond this
distance contribute negligibly to shielding. 

Strictly speaking, this method should only provide an accurate
estimate of $N_{\rm eff}$ if the velocity gradient is both large and
monotonic and the intrinsic line widths are negligible in comparison
to the Doppler cores. If instead the Lorentz wings (and line overlap)
become important, these will render the Doppler shifts irrelevant and
this method will underestimate self--shielding as a
result. None the less, it may prove useful, particularly if motions of
the fluid are supersonically turbulent.  In fact, a similar method
has been fruitfully employed in the analogous case of the escape
fraction of photons from dense, metal--free gas
\citep[e.g.][]{Yosh+06}.

In order to generalize to a non--spherically symmetric geometry, one
option is to define a single $L_{\rm Sob}$ from the mean Sobolev
length over all directions, which we will refer to as $L_{\rm Sob}^s$.
Alternatively, in direct analogy with the three-dimensional
self-shielding calculation, we could find $L_{\rm Sob}$ and the
corresponding dissociation rate for each sightline and take the mean
rate over all directions (hereafter $L^k_{\rm Sob}$). Clearly, this
approach should provide a reasonably accurate estimate of the
dissociation rate if the corresponding column density is a good
approximation for $N_{\rm eff}$ along every direction. Finally, we
could take the minimum Sobolev length ($L^{\rm min}_{\rm Sob}$), or an
average column density ($L^N_{\rm Sob}$), over all directions.

\noindent{\it A ``Sobolev-Like'' Length} \\
A method akin to the Sobolev length has been recently employed and shown 
to provide a fairly accurate estimate of the integrated column density by 
\citet{GTK09}, though in a different context than that of the present work 
(they model self--shielding in individual star--forming regions of a Milky 
Way progenitor). In this case, a characteristic length is obtained from 
\begin{equation}
L'_{\rm Sob} \equiv \frac{\rho}{|\nabla\rho|},
\end{equation}
thereby defining a distance over which the gas density, $\rho$, should
be significantly diminished, and assuming that the optical depth beyond 
$L'_{\rm Sob}$ is negligible. 

We expect this method to prove most useful when the length scale of
significant decrease in the gas density is shorter than those over
which large variations in the ${\rm H_2}$ fraction or the gas velocity
occur. Note that the aim is to define a distance beyond which the gas
contributes negligibly to shielding, so extrapolation along lines of
sight for which ${\rm d\rho/ ds} > 0$ is not meaningful (this happens
for off-centre points, in directions toward the halo centre).  Thus,
we define $L^{'s}_{\rm Sob}$, $L^{'k}_{\rm Sob}$, and $L^{'N}_{\rm Sob}$
in a similar manner as described above for the traditional Sobolev--like 
length, but include in the averaging only those sight--lines for which 
${\rm d\rho/ ds} < 0$ . \\

\noindent{\it The ``Six--Ray Approximation''} \\ Finally we consider
one non--local method, variations of which have been implemented in
simulations by, e.g., \citet{Yosh+03,Yosh+07} and
\citet{GML07a,GML07b}. In this case, the exact $N_{\rm H_2}$ is
obtained by integrating the ${\rm H_2}$ profile along six lines of
sight parallel to the Cartesian axes. The value of the shield factor
for each is obtained from one of the DB96 fits (equations
\ref{eq:DB36} or \ref{eq:DB37} below) and the final rate is found by
averaging over six directions.\footnote{Note that \citet{Yosh+03} use
instead the shield factor calculated for the line of sight with
minimum ${\rm H_2}$ column density.}  In spite of the low angular
resolution, this approach is expected to be reasonably accurate unless
the gas is (supersonically) turbulent, in which case neglecting
Doppler shifts of the LW lines will likely cause the integrated
$N_{\rm H2}$ to substantially exceed the effective self-shielding
column density. 
\begin{figure*}
\includegraphics[clip=true,trim=0in 3.8in 0in 0.4in,
height = 2in,width = 4.8in]{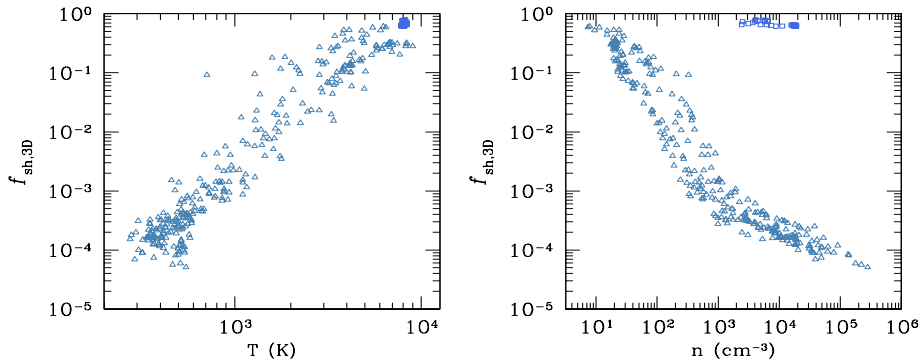}
\caption{The numerical results for the shield factor in three--dimensions, 
$f_{\rm sh,3D}$, are shown at the local temperature (left--hand panel) and 
particle density (right--hand panel). Points are included from all five 
halo outputs; those from cold and hot haloes are denoted (here and in all 
subsequent figures) by open triangles and squares, respectively.}
\label{fig:3Dresults}
\end{figure*}
\subsubsection{Analytic approximations for the photodissociation rate}
\label{sec:DB96}

Self--shielding by ${\rm H_2}$ has received attention over the years
for its importance in the context of interstellar clouds
(e.g. \citealt{HWS71,Shull78,FGK79,dJBD80,Abgrall+92,Heck+92,
LeBourlot+93};~DB96). A number of analytic models for the attenuation
of the incident flux have been put forward, as explicit calculations
of the radiation field as a function of both cloud depth and frequency
have not previously been feasible.  In the last decade, however, studies 
have most often employed the expressions provided by DB96.
These authors model a semi--infinite, static slab of gas irradiated on
one surface, and parametrize the dissociation rate with a ``shield
factor'' as:
\begin{equation}
k_{\rm diss}(N_{\rm H_2},T) = f_{\rm sh}(N_{\rm H_2},T) \times
k_{\rm diss}(N_{\rm H_2} = 0,T),
\end{equation}
where $k_{\rm diss}(N_{\rm H_2} = 0)$ is the optically--thin rate. 
They show that, at low temperatures ($T \sim$ a few $\times 10^2$ K), 
suppression of the optically--thin rate can be well approximated by 
a simple power-law that depends only on the ${\rm H_2}$ column 
density as: 
\begin{equation}
f_{\rm sh,36} \left(N_{\rm H_2}\right) = {\rm min}
\left[1,\left(\frac{N_{\rm H_2}}{10^{14}~{\rm cm^{-2}}}
\right)^{-3/4}\right].
\label{eq:DB36} 
\end{equation}
These authors also provide a slightly more complicated functional
form, which attempts to incorporate a temperature dependence due to
thermal broadening of the lines, and which fits their results more
accurately:
\begin{multline}
f_{\rm sh,37}\left(N_{\rm H_2}, T\right) =
\frac{0.965}{\left(1 + x/b_5\right)^2}
+ \frac{0.035}{\left(1 +  x\right)^{0.5}}\\
\times \exp\left[-8.5 \times 10^{-4} \left( 1 +  x\right)^{0.5}\right].
\label{eq:DB37}
\end{multline}
Here $x \equiv N_{\rm H_2}/ 5 \times 10^{14}~{\rm cm^{-2}}$, $b_5
\equiv b/10^5~{\rm cm~s^{-1}}$, and $b$ is the Doppler broadening
parameter (equations \ref{eq:DB36} and \ref{eq:DB37} are given in DB96
as their equations 36 and 37, respectively, as indicated by the subscripts). 
Because \ref{eq:DB37} is the more accurate of the two, this will be 
the focus of our discussion henceforth.

These expressions have been ubiquitously used to model self-shielding
in simulations. While it is often noted that they are only strictly
valid in the static limit, it should be emphasized that DB96 make
several assumptions in their modeling which limits the applicability
of the expressions they derived. Most importantly, they consider a
${\it cold}$ gas, with $T \lsim {\rm~a~few} \times 10^2$
K. Furthermore, they assume either (i) a steady--state rovibrational
distribution that includes absorptions from $v > 0$ or (ii) an
isothermal gas with ${\rm H_2}$ level populations given by a Boltzmann
distribution. Also included are shielding by dust and the formation of
${\rm H_2}$ on grains.  None the less, their fitting formulae are
often implemented in contexts quite different from the original study,
in which these assumptions are not well motivated. In light of this,
\S~\ref{sec:Results} addresses the accuracy of
equation~(\ref{eq:DB37}) in the present context, i.e. for metal--free
gas with a wide range of densities $10 \lsim n_{\rm tot} \lsim
10^{6}~{\rm cm^{-3}}$ and temperatures $300 \lsim T \lsim 10^4$K, and
in which dust, as well as UV--pumping from excited vibrational states
is likely negligible (see \S~\ref{sec:LevelPopulations}).

\section{Results}
\label{sec:Results}
\begin{figure*}
\includegraphics[clip=true,trim=0in 0in 0in 0.4in,height = 4.in,width = 4.8in]{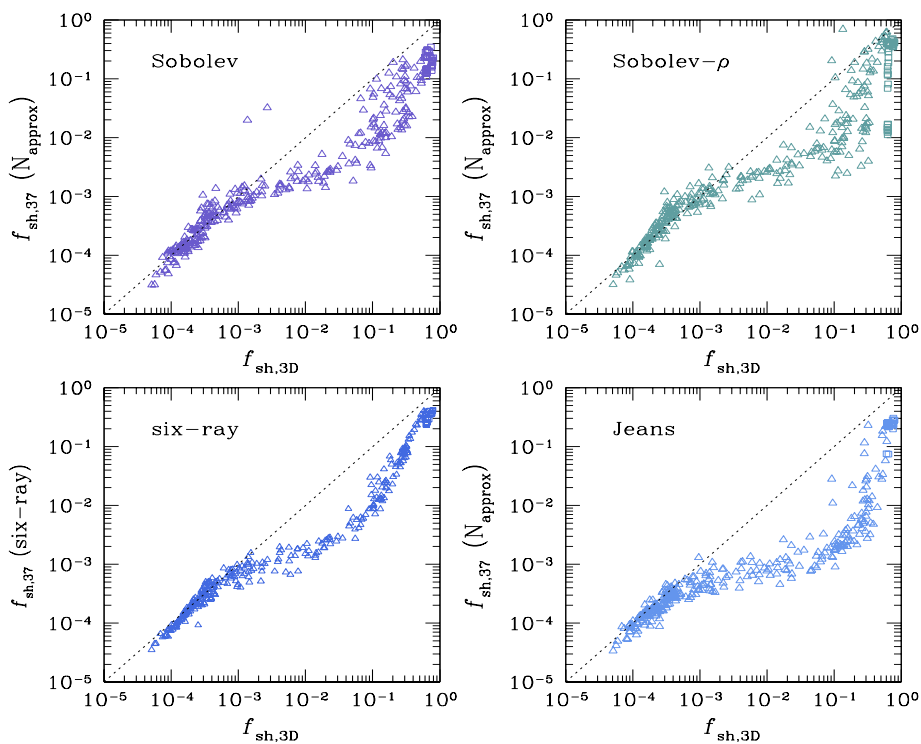}
\caption{Comparison of our numerical results for the average shield
  factor in three dimensions, $f_{\rm sh,3D}$, with those given by the
  analytic fitting formula ($f_{\rm sh,37}$ in equation~(\ref{eq:DB37}), 
  taken from DB96). Clockwise from the upper left, the column density 
  is specified by the best--fitting Sobolev and Sobolev--like (``Sobolev--$\rho$'') 
  methods, ($L^k_{\rm Sob}$ and $L^{'s}_{\rm Sob}$ 
  respectively; see \S~\ref{sec:Napprox}), and the Jeans length. The
  lower left panel shows results of the only previous non--local
  approach, the ``six-ray method,'' in which equation~(\ref{eq:DB37})
  is used to calculate the shield factor along six sight lines, and an
  average over these gives $f_{\rm sh,37}$(six--ray). Non--local
  effects of velocity and temperature gradients are included in the
  calculation of $f_{\rm sh,3D}$, as described in \S~\ref{sec:3Dshielding}.}
\label{fig:DB+Napprox}
\end{figure*}
The numerical results for the photodissociation rate in
three--dimensions, parametrized (as above) by a shield factor,
$f_{\rm sh,3D}$, are shown in Figure \ref{fig:3Dresults} as a function
of the local temperature and particle density of each point. The trend
toward decreased optical depth at high temperatures and low densities
is due to the characteristic structure in these haloes, in which
regions farther from the dense core are both hotter and more rarefied,
and typically see lower shielding columns (see
Figure~\ref{fig:profiles}). The exception is the hot haloes, in which
the gas remains optically thin even at small radii, where $n \gsim
10^{3-4}~{\rm cm^{-3}}$, due to the low molecular fraction. As a
result, these hot haloes (outputs 3 and 4) will not be further
addressed in great detail, as they add little to the discussion of
self--shielding.

Figure \ref{fig:DB+Napprox} shows these results in comparison with
several approximate methods. The four panels show $f_{\rm sh}$ given
by equation~(\ref{eq:DB37}), and in each, the column density is
specified by one of the methods discussed in \S~\ref{sec:Napprox}.
Because studies typically must employ estimates for $N_{\rm H_2}$ as
well as an analytic fit for the shield factor, the overall
discrepancies seen here are representative of what would be realized
in a simulation.

A clear trend is immediately apparent in Figure \ref{fig:DB+Napprox},
which appears independent of how the column density is estimated: all
approximate methods significantly overestimate shielding compared to
our numerical results in the range $10^{-3} \lsim f_{\rm sh,3D} \lsim
0.3$, while each is considerably more accurate at smaller values of
$f_{\rm sh,3D}$. In principle, this discrepancy may arise from
physical effects, i.e. Doppler shifts in the LW absorption lines,
causing self--shielding to be weaker than in a static gas, or
variations in the gas temperature, resulting in depopulation of states
that contribute most to dissociation.  Alternatively, it may be due to
inaccuracy of the analytic fit for the $f_{\rm sh}$ itself. In what
follows, we examine each of these possibilities in turn, to determine
how much each effect contributes to the observed trend.

\subsection{Analytic approximations for ${\bf {\it f}_{sh}}$}
\label{sec:FittingFormulae}

\begin{figure*}
\includegraphics[clip=true,trim=0in 3.8in 0in 0.4in,height = 2in,width = 4.8in]{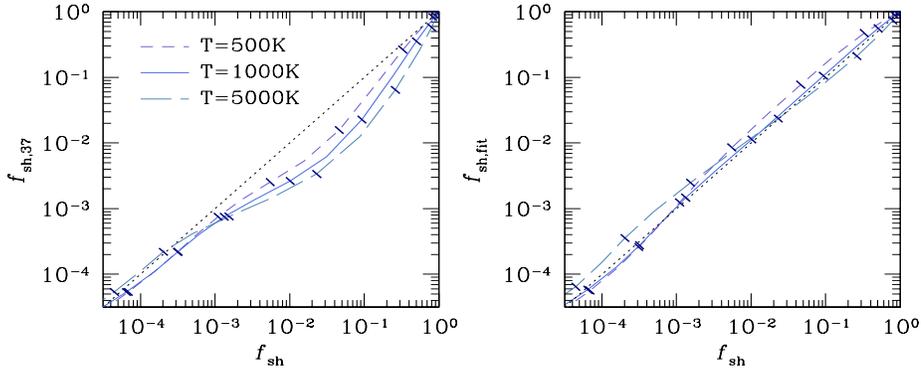}
\caption{The results from equations~(\ref{eq:DB37}; left panel) and
  (\ref{eq:newfit}, with $\alpha = 1.1$; right panel) are compared against
  the numerical results for $f_{\rm sh}$ at $T=$ 500, 1000, and 5000K.  
  In the latter, the gas is treated as an isothermal and static slab,
  irradiated uniformly on one side. Tick marks along each curve
  indicate ten--fold decreases (left to right) in the column density,
  with $N_{\rm H_2} = 10^{20}{\rm~cm^{-2}}$ at the lower left; the
  dotted curves are 45-degree lines for reference.}
\label{fig:FittingFormulae}
\end{figure*}

The left panel of Figure~\ref{fig:FittingFormulae} shows a comparison 
of the shield factor obtained from the DB96 expression 
(equation~\ref{eq:DB37}), against the results from our numerical 
calculations at $T =$ 500, 1000, and 5000K. In the latter, the gas 
is modeled as a static and isothermal slab, in order to isolate the 
accuracy of equation~(\ref{eq:DB37}) from the effects of temperature 
and velocity gradients.  It is apparent from this figure that the trend of
discrepancies seen in all four panels of Figure \ref{fig:DB+Napprox}
is caused in large part by the inaccuracy of the fitting formula
itself at temperatures greater than a few hundred Kelvin. (Note that,
as shown in Figure~\ref{fig:3Dresults}, all points in the simulation 
with $f_{\rm sh,3D}\gsim 10^{-3}$, which are discrepant, are at $T >500$ K.)

We have found that the large discrepancies -- equation~(\ref{eq:DB37})
underestimates the numerical results by up to an order of magnitude --
are due to the temperature dependence of $f_{\rm sh}$ for thermalized
${\rm H_2}$ populations. To illustrate this, consider a gas in local
thermodynamic equilibrium (LTE) at a (uniform) temperature of a few
hundred Kelvin, so that only the lowest rotational states within the
vibrational ground state will be significantly populated (for
reference, the energies of the $v=0,~J=1,2$ states are $\approx
170,509$ K). As the temperature is increased, the populations will be
diluted over a greater number of rotational levels (e.g. at $T \sim$
several thousand Kelvin, non--negligible populations build up in $J
\lsim 15$). The upshot of thus spreading absorbers over a greater
number of states is that shielding becomes weaker\footnote{The
importance of diluting the populations over higher energy states was
pointed out by \citet{FGK79}, who showed that for a thermal
distribution, including absorption from $J = 2$ changed their results
by $\sim$ 40 per cent compared to those in ortho and para ground states were
populated.}. This effect is not modeled by the DB96 expression
(equation~\ref{eq:DB37}), which accounts only for decreased shielding
due to thermal broadening of the lines. As a result, we find that it
is accurate only for a rotationally--cold gas, i.e. when molecules
occupy only the first few $J$ states (we show this explicitly in
Figure \ref{fig:groundstate}).

The inaccuracy of this analytic fit at high temperatures has been
noted previously by \citet{AS07}, who proposed that it may be remedied
by artificially increasing the thermal broadening parameter used in
equation~(\ref{eq:DB37}). We have found that this method yields very
little improvement in accuracy; however, a much better fit to our
numerical results is obtained with only a slight modification of
equation~(\ref{eq:DB37}) as follows. We treat $\alpha$, below, as a
free parameter, where $\alpha = 2$ in the original expression:
\begin{multline}
f_{\rm sh}\left(N_{\rm H_2}, T\right) =
\frac{0.965}{\left(1 + x/b_5\right)^\alpha}
+ \frac{0.035}{\left(1 +  x\right)^{0.5}}\\
\times \exp\left[-8.5 \times 10^{-4} \left( 1 +  x\right)^{0.5}\right].
\label{eq:newfit}
\end{multline}
We find that $\alpha = 1.1$ improves the fit drastically in the high
temperature regime, with little accuracy sacrificed at low
temperatures. Physically, this result makes sense, since reducing
$\alpha$ weakens the temperature effect, which, as we argued, is
overestimated in the original equation.  The modified expression
agrees with the numerical results to within a factor of two at $500< T
<5000$ K, and $N_{\rm H_2} \leq 10^{20} {\rm~cm^{-2}}$, as shown by
the solid curves in Figure \ref{fig:FittingFormulae}. Furthermore, as
the figure shows, the largest discrepancies occur (a) at low
temperature, $T = 500$ K, and larger values of the shield factor,
$f_{\rm sh} \gsim 10^{-2}$, or (b) high temperature, $T \gsim 5000$ K
and $f_{\rm sh} \lsim 10^{-3}$. In the context of our study, these
combinations are, in fact, never physically realized, as low
temperatures ($T \approx 500$ K) are only reached via ${\rm H_2}$
radiative cooling, implying that in cold gas, large column densities
will have built up and the shield factor will be small; for the same
same reason, the gas cannot remain hot once it has become strongly
self--shielding (see the temperature and density profiles in
Figure~\ref{fig:profiles}). Overall, we find that the agreement
between equation~(\ref{eq:newfit}), with $\alpha = 1.1$, and the
results for $f_{\rm sh,3D}$ in the haloes is $\approx 15$ per cent.

\subsection{Impact of velocity gradients}
\label{sec:vgradients}

As previously described, the existence of internal velocity gradients
will cause a given parcel of gas to see more flux than it would in a
static gas, owing to Doppler shifts of the LW resonances in regions of
the fluid moving at large relative velocities. This can lead the
actual $N_{\rm H_2}$ to exceed the effective column density by a large
factor, and could thus also contribute to $f_{\rm sh}(N_{\rm approx})$
overestimating shielding compared to the numerical results (see Figure
\ref{fig:DB+Napprox}).

\begin{figure*}
\includegraphics[clip=true,trim=0in 5in 0in 0.43in, height=2in,width=6.9in]{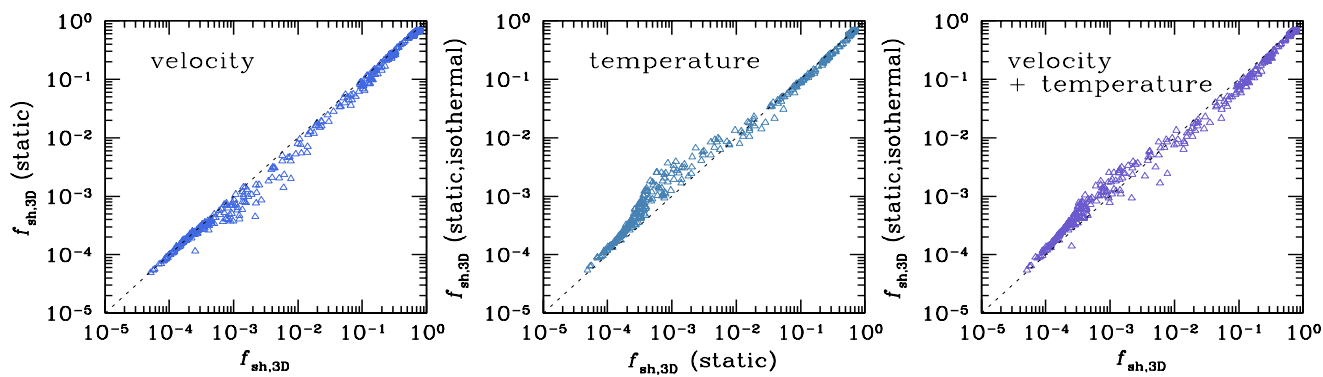}
\caption{{\it Left--hand panel:} the fiducial numerical results,
      $f_{\rm sh,3D}$, are compared to those in which the gas is
      assumed to be static.  {\it Middle panel:} isolates the effect
      of setting the temperature constant along each sightline (with
      the gas again assumed to be static along both axes).  {\it
      Right--hand panel:} shows the combined effect of the velocity
      and temperature variations, by comparing the fiducial results
      with those in which the gas is assumed to be both isothermal and
      static.  Note that, here, ``isothermal'' does not imply that
      temperature is the same for all points, but rather that $T =
      T_0$ along all sight lines, where $T_0$ is the temperature where
      the dissociation rate is calculated.}
\label{fig:isothermal+static} 
\end{figure*}

To quantify the magnitude of this effect, we repeated our
three-dimensional calculations with gas velocities artificially set to
zero everywhere.  The results, shown in the left panel of Figure
\ref{fig:isothermal+static}, can be divided roughly into three
regimes: strongest, intermediate, and weakest shielding; the
transition between the first (latter) two occurring at $f_{\rm sh,3D}
\simeq 4\times 10^{-4}$($2 \times 10^{-2}$). The dissociation rate is
increased in the non--static case by an average of 6, 65, and
20 per cent, respectively, in the three regimes (though in the intermediate
case, certain points see dissociation rates that are factors of up to
4-5 times greater due to relative gas velocities).

\begin{figure}
    \includegraphics[clip=true,trim=0.2in 4in 3.65in 0.1in, 
height=2.7in,width = 3.4in]{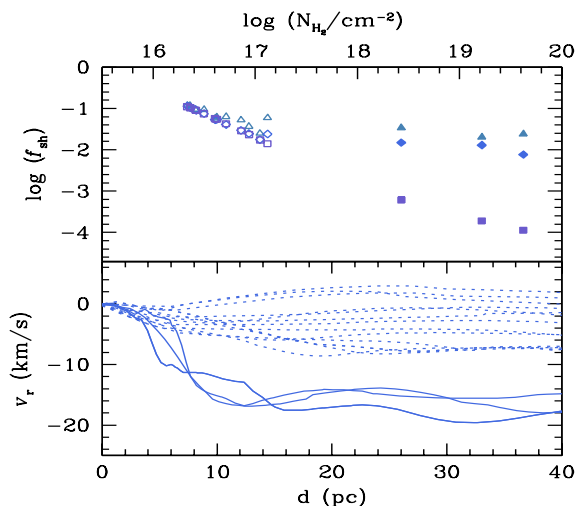}
    \caption{{\it Upper panel:} comparison of the
numerically--calculated shield factors along individual sight lines
from a single point (triangles) with those obtained when the gas
is assumed to be static (diamonds) or both static and
isothermal (squares). The full spherically averaged photodissociation 
rate is increased 23 per cent (30 per cent) compared to the static
(static and isothermal) case. {\it Lower panel:} relative line of
sight velocities along the same directions as shown in the upper
panel.  The three lower (solid) curves correspond to the three sight 
lines with largest total $N_{\rm H_2}$ (closed symbols) in the upper panel.}
\label{fig:sightlines}
\end{figure}

In order to understand why this effect operates differently in the
three regimes, let us first consider points where shielding is
weakest, $f_{\rm sh,3D} \gsim 2 \times 10^{-2}$. In Figure
\ref{fig:sightlines}, the shield factors are shown for individual
sight lines emanating from a single point in this regime; these are
shown for both the static and non--static cases (closed and open
triangles, respectively) at the integrated $N_{\rm H_2}$ in each
direction. In the lower panel, the relative line of sight velocities
along all 16 directions are shown.  Three directions are distinct from
the rest, in that they see the largest velocities as well as the
greatest integrated column densities -- the three bottom curves in the
lower panel correspond to the three right--most points in the upper
panel. This highlights an important point about the dynamics in these
simulations: a coherent flow of the fluid onto the densest region
dominates the gas velocities, rather than turbulence.  Thus, large
frequency shifts in the ${\rm H_2}$ lines occur only along sight lines
aligned with this flow, and while they clearly do change the
dissociation rate in these directions, variations in these smallest
values of the shield factor have little effect on the spherically
averaged value.  In the strongest shielding regime, the column
densities are sufficiently high that large damping wings of the lines
render frequency shifts irrelevant. We have verified this by
artificially decreasing the column density of each slab along these
sightlines by a factor of 100 and repeating the full calculations with
velocities both on and off; in this test, $f_{\rm sh,3D}$ becomes
several times larger in the non--static case.

In the intermediate regime, the damping wings of the lines are not
significant, and frequency shifts have a larger impact on the
dissociation rate as a result. This is ${\it not}$, however, because
the total column densities are lower than in the previous case.  The
integrated $N_{\rm H_2}$'s along sight lines emanating from these
points are, in fact, comparable to those in the strongest shielding
regime, but the points themselves are in regions with much higher
temperatures (i.e. $T \gsim 10^3$K, compared to a few hundred Kelvin;
see Figure \ref{fig:3Dresults}). This highlights the importance of
coupling between the effects of temperature and velocity gradients. As
previously discussed, increased temperature dilutes the ${\rm H_2}$
populations over a greater number of rovibrational levels; the upshot
here is that diminished Lorentz wings of transitions originating from
these no longer cancel the effects of Doppler shifts. In other words,
the column density in a given rovibrational level {\it is} smaller
than in the strongest shielding regime, though the total $N_{\rm H_2}$
for individual sight lines is comparable.

\subsection{Impact of temperature gradients}
\label{sec:Tgradients}

While not as often addressed as the effect of frequency shifts,
variations in the gas temperature along a given line of sight may also
significantly alter the self--shielding behaviour, particularly for 
a gas in LTE. There are two distinct temperature-induced effects: 
(i) changes in the level populations (specific to the LTE case), 
and (ii) changes in the widths of individual lines.

Consider the population effect first. For a given parcel of gas with
temperature $T_0$, along a given direction, deviations in the
temperature from $T_0$ and the resulting change in the rotational 
distribution may either increase or decrease the populations 
of the levels from which most dissociating transitions originate. 
In general, if (a) $T_0$ is small ($\sim$ a few $\times 100$K) and 
the temperature is increased along a sightline, or (b) the 
{\it difference} between the temperature of the shielding gas and 
$T_0$ is very large, the result is to depopulate the states from 
which most UV--pumping occurs, and thus to decrease the effective 
column density. On the other hand, if the temperature variations 
along a sightline are not large, the shielding may be increased
compared to the isothermal case, owing to boosted populations in states 
from which the strongest transitions (i.e. those for which the product 
of the oscillator factor and dissociation probability is largest) originate. 

Interestingly, we find that the line-width effect may be equally or
more important than the above population--induced changes. Consider
again our gas parcel with temperature $T_0$, shielded, for simplicity,
by a single slab of gas at $T < T_0$. The narrowed thermal widths of
absorption lines in the slab result in overall much weaker shielding,
in comparison to the isothermal case (though the opacity at the line
centre is actually increased). Furthermore, because this effect may
act in the same direction as the population--induced changes,
shielding can be greatly reduced for the case in which $T < T_0$. This
is illustrated in Figure \ref{fig:sightlines}, in which the open
squares show the results when both the velocity and temperature are
held constant along all sight lines. Recall that these sight lines
emanate from a point which lies in the weakest shielding regime; the
temperature is $\sim 3500$K where the dissociation rate is calculated,
while along the rays passing through the dense central core of the
halo (for which $N_{\rm H_2}$ is largest), the temperature drops as
low as $300-400$K. The resulting narrowing of the thermal cores in the
shielding gas, along with depopulation of excited rotational states
from which a large fraction of the dissociating transitions occur,
drastically reduces the effective column density and correspondingly
increases $f_{\rm sh}$. In principle, this effect could also
contribute to the approximate methods overestimating self--shielding
as compared to the numerical results; however, these changes again are
typically only significant for the smallest values of the shield
factor (sightlines which are directed through the dense core), and
thus have a small effect on the spherically averaged value.

Finally, we consider the simplified model of a single shielding slab
with $T > T_0$. In this case, the larger thermal widths in the
shielding gas yield greater optical depth in the wings of the Voigt
profiles of the dissociating transitions, and this can outweigh the
reduced line centre opacity. This effect depends on $N_{\rm H_2}$
along the sightline in question, and there is a column
density--dependent $T_{\rm max} > T_0$, at which $f_{\rm sh}$ is a
minimum, above which further increases in the temperature lead to
weaker shielding. This is not important in the weakest shielding
regime, as $T_0$ here is already very high (see Figure
\ref{fig:3Dresults}), so that few sight lines see a significant
increase in the temperature. In the strongest shielding regime, the
broader thermal cores are unimportant because the absorption lines
already have large damping wings, so that extremely high temperatures
would be required for this effect to operate. It is, however,
important in the intermediate regime; as illustrated in the middle
panel of Figure \ref{fig:isothermal+static}. This panel shows that
shielding is always {\it weaker} when the temperature is held constant
along sight lines for $f{\rm _{sh,3D}(static)}< 10^{-2}$.  (Here, gas
velocities have again been set to zero so as to isolate the effect of
non--uniform temperature, which couples to that of velocity
gradients). However, this cannot be entirely attributed to the
line--width effect, as temperature changes in this regime often also
induce stronger shielding owing to increased populations of important
lines, as mentioned above. Interestingly, the cumulative effect
counteracts that of Doppler shifts, which are also more important in
this regime, with the result that ``turning on'' both temperature and
velocity gradients causes only a small scatter around the original
results, as illustrated in the right--hand panel of Figure
\ref{fig:isothermal+static}.

\begin{figure*}
\includegraphics[clip=true,trim=0in 0in 0in 0.4in,height=4in,width = 4.8in]{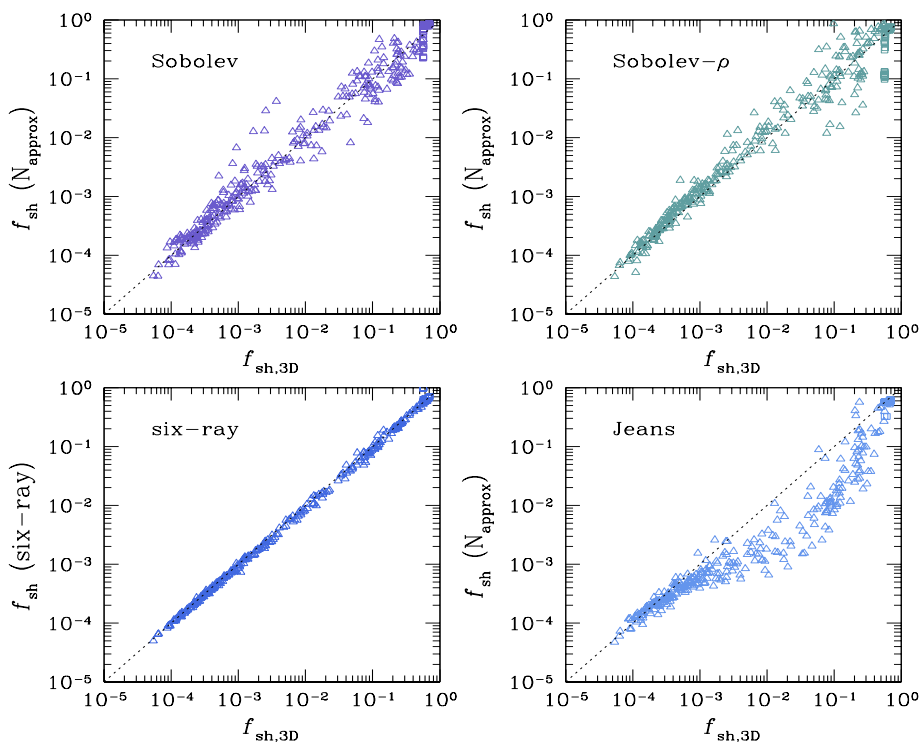}
    \caption{Comparison of the numerical results for the shield factor
      in three dimensions, $f_{\rm sh,3D}$ to those obtained from a
      single slab model with (locally) estimated column densities,
      $N_{\rm approx}$. In the former, velocity and temperature are held
      constant along all sightlines, in order to isolate the accuracy of
      each $N_{\rm approx}$ method. Clockwise from the upper left, 
      $N_{\rm approx}$ is specified by the best--fitting Sobolev length 
      and Sobolev-like (``Sobolev--$\rho$'') length ($L^k_{\rm Sob}$ and 
      $L^{'s}_{\rm Sob}$, respectively; see \S~\ref{sec:Napprox}), and 
      the Jeans length. The lower left panel shows the results of the only
      non--local approach, the ``six--ray method,'' in which the
      shield factor is calculated for six sight lines using the
      integrated $N_{\rm H_2}$ and an average over these gives 
      $f_{\rm sh}$(six--ray). The overall fractional errors of each 
      $f_{\rm sh}(N_{\rm approx})$ are shown in Table 2.}
\label{fig:Napprox}
\end{figure*}

\subsection{Approximate self--shielding column densities} 

The only remaining source of discrepancy between the numerical and
approximate results for $f_{\rm sh}$ are the column density
approximations, $N_{\rm approx}$, obtained either from local
quantities, or from the six--ray method.  In order to isolate the
accuracy of each method described in \S~\ref{sec:Napprox}, we have
calculated the dissociation rate numerically, assuming a single slab 
of shielding gas with $N_{\rm H_2} = N_{\rm approx}$. The results are
compared to the exact $f_{\rm sh,3D}$, in Figure \ref{fig:Napprox}, 
and the mean errors from each method are quantified in Table 2.
In the three--dimensional calculation, the gas temperature and
velocity are held constant along all sightlines, so as to isolate the
accuracy of each method for specifying $N_{\rm approx}$
\footnote{We note that it is somewhat awkward to compare the results 
from the Sobolev length to the numerical results assuming a static slab --
formally the Sobolev length is infinite in static gas. However, Figure 
\ref{fig:Napprox} clearly illustrates that the efficacy of this method 
here {\it cannot} be attributed to the presence of velocity gradients.}.

Over the full range of radii we consider, the Sobolev length proves
the most accurate purely local estimate for the effective column
density. This is perhaps surprising, given the results in
\S~\ref{sec:vgradients}, where we have shown that velocity gradients
do not alter the observed LW flux within these haloes
significantly. In fact, because of the unimportance of velocity
gradients for our results, we might expect that the Sobolev length
should not provide a meaningful length scale, $L_{\rm char}$, for
self--shielding.  It is worth noting that this method is also the only
one considered here, to the best of our knowledge, which has not been
implemented in simulations in the context of self--shielding (though
it is commonly used in the opposite case, to compute the escape
fraction of photons traveling outward).  Indeed, while it is not
obvious how to justify the success of this method, its potential
usefulness in simulations should be highlighted; though the scatter is
larger than from the six--ray method, the computational expense is
dramatically reduced.

In fact, Figure \ref{fig:Napprox} shows that, for $f_{\rm sh} \lsim
10^{-3}$, all three local methods provide surprisingly accurate
estimates of the column density, given the relatively crude
assumptions made in approximating the characteristic length scale for
shielding. However, while both Sobolev and Sobolev--like methods are
also reasonably accurate in the less shielded regime, in agreement
with SBH10, we that the Jeans length typically underestimates the
dissociation rate by an order of magnitude or more at $f_{\rm sh,3D}
\gsim 10^{-3}$, i.e. in regions where the number density is low,
$n_{\rm tot} \lsim 10^{4}~{\rm cm^{-3}}$, and temperature is high, $T
\gsim 10^3$K as shown by Figure \ref{fig:3Dresults} (see also Figure 9
in SBH10).

Not surprisingly, the six--ray method proves more accurate than the
``local'' methods. In fact, in the absence of temperature and velocity 
gradients, six sightlines would suffice for convergence of the spherically 
averaged rate. However, larger discrepancies occur if the six--ray 
results are compared to the full three--dimensional calculation (this comparison
is not shown) rather than the static and isothermal case; this is simply
because the integrated $N_{\rm H_2}$ is then, in general, not equal to the 
effective self--shielding column density. In this case, the average ratio 
($1 \sigma$ scatter) for  $f_{\rm sh}$(six--ray)/$f_{\rm sh,3D}$ = 1.2 (0.1) 
and 1.1 (0.5) for points with particle densities above and below $10^4 
~{\rm cm^{-3}}$, respectively (c.f. Table 2). Recall that most often this 
method is implemented by integrating to find the total $N_{\rm H_2}$ along 
each of the directions and plugging this into equation~(\ref{eq:DB36}), 
or (\ref{eq:DB37}); therefore, it does not account for variations in the 
temperature and velocity along the rays. (An exception is the study by 
\citealt{Yosh+07}, in which $N_{\rm H_2}$ is summed along each ray, 
excluding gas particles with relative velocities significantly larger 
than the local thermal velocity.) 

\begin{table}
\label{tbl:errors}
\caption{The accuracy of each method for estimating the column density 
  (see \S~\ref{sec:approximations}) is quantified by the mean of  
  $f_{\rm sh}(N_{\rm approx})/f_{\rm sh,3D}$(isothermal, static).
  Ratios are reported separately for regions with particle densities 
  above and below $10^4 ~ {\rm cm^{-3}}$, corresponding roughly to a 
  cut--off around $f_{\rm sh,3D} \approx 10^{-3}$. Numbers in parentheses 
  indicate the $1\sigma$ scatter. Boldface values highlight the best--fitting 
  Sobolev and Sobolev--like ($L'_{\rm Sob}$) methods.}
\begin{center}
\setlength{\extrarowheight}{0.2cm}
\begin{tabular}{l c c}
\hline\hline

& $n < 10^4 ({\rm cm^{-3}})$ & $n > 10^4 ({\rm cm^{-3}})$ \\
\hline
$L_{\rm Jeans}$ & 0.50 (0.38) & 0.96 (0.17)\\

$L^s_{\rm Sob}$ & 1.7  (4.3)  &  1.1 (0.3)\\

$^{a.} L^k_{\rm Sob}$ & {\bf 1.3 (1.3)} & {\bf 1.0 (0.3)}\\

$L_{\rm Sob}^N$ & 0.48 (0.44) & 0.46 (0.23) \\

$L_{\rm Sob}^{\rm min}$ & 6.1 (25) & 2.1 (0.7)\\

$^{b.} L^{'s}_{\rm Sob}$ & {\bf 1.4 (0.9)} & {\bf 1.1 (0.2)}\\

$L^{'k}_{\rm Sob}$ &  0.66 (0.75) &  0.61 (0.14) \\

$L^{'N}_{\rm Sob}$ &  0.78 (0.49) & 0.67 (0.31)\\

$L^{\rm 'min}_{\rm Sob}$ & 3.0 (6.5) & 1.6 (0.3)\\

six-ray & 0.93 (0.09) & 0.97 (0.05) \\
\hline\hline
\end{tabular}

\begin{tabular}{p{8.2 cm}}
$^{a.}$ Recall that, in this method, the Sobolev 
length for each sightline is used to calculate the dissociation 
rate, and the final rate is obtained from a simple average over 
all directions.\\
$^{b.}$ In this case, the rate is calculated given a single column 
density specified by the spherically averaged ``Sobolev--like'' 
length.\\
\end{tabular}

\end{center}
\end{table}

\subsection{Angular resolution for ${\bf {\it f}_{sh}}$ in 3D}
\label{sec:ConvergenceTest}

In order to determine the required angular resolution for the
numerical calculations, we performed a convergence test for the
spherically averaged $f_{\rm sh,3D}$ using 50 points selected from the
five simulation outputs, spanning the full range of radii we consider,
$0.1 \lsim R \lsim 10$pc.  The mean error in the shield factor,
$f_{\rm sh,3D}$, for these points was found to be three (two) per cent
for an increase from $16 \to 25$ sightlines ($25 \to 49$).  In
comparison, the mean error for an increase from six to 16 sightlines
was found to be $\sim 9$ per cent; note that, as in the ``six--ray''
method described above, the six sightlines were aligned parallel with
the Cartesian axes, but in this case, the full calculation was
performed -- i.e.  including density, temperature, and velocity
gradients. All of our calculations for $f_{\rm sh,3D}$ are therefore
obtained from averaging over 16 sight lines, which sample evenly in
the azimuthal angle and in the cosine of the polar angle.

An additional convergence test was performed to determine the required
angular resolution for the mean Sobolev and Sobolev--like lengths
($L^s{\rm Sob}, L^{'s}_{\rm Sob}$). For 25 points selected, again,
over the full range of radii considered, the mean error in the
spherically averaged Sobolev and Sobolev--like lengths are 15 and 17
per cent for an increase from $16 \to 25$ sightlines, 10 and 11 per
cent ($25 \to 49$), and 1 and 3 per cent ($49 \to 100$), respectively.
As a result, 49 sightlines were used in all calculations for the
Sobolev and Sobolev-like methods. (Note that these specific numbers
are determined by the tiling method we use, which always results in a
perfect square for the number of sightlines.)

\section{Implications and Caveats} 
\label{sec:Implications}

\subsection{Cooling and collapse in simulated haloes} 
\label{sec:jcrit}

\begin{figure}
  \includegraphics[clip=true,trim=0.2in 4in 3.65in 0.4in,                    
    height = 2.5in,width = 3.4in]{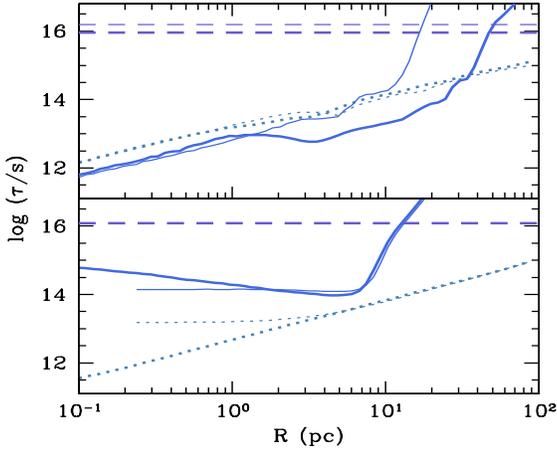}                                            
    \caption{{\it Upper panel:} spherically averaged ${\rm
        H_2}$--cooling (solid curves) and dynamical (dotted curves)
      time--scales are shown for the cold haloes; the Hubble time for
      each output is shown by horizontal (dashed) lines, with the same
      cosmological parameter choices as in SBH10, ($\Omega_{\rm m},
      \Omega_\Lambda, h$) = (0.279, 0.721,0.701). Thick and thin
      curves denote the results for outputs 1 and 5,
      respectively. (The dynamical time-scales are nearly identical
      for these outputs, making the thick and thin curves
      indistinguishable.) The results for output 2 are nearly
      identical to those of output 1, and have been omitted for
      clarity.  {\it Lower panel:} same as above for the hot haloes,
      with thick (thin) curves denoting the time-scales for output 3
      (4).}
  \label{fig:timescales}
\end{figure}

In the preceding section, we have shown that common approaches to
model self--shielding in simulations can introduce inaccuracies of
over an order of magnitude in the ${\rm H_2}$ dissociation rate,
depending on the level of the shielding and on the method being
used. Because the cooling properties of metal--free gas below $10^4$ K
depend sensitively on the ${\rm H_2}$ abundance, errors in the
dissociation rate can significantly affect the thermal and dynamical
histories of gas in haloes under UV irradiation. In order to quantify
such errors, and to specify {\it when} inaccurate values of $f_{\rm
  sh}$ will have the greatest impact, we show in Figure
\ref{fig:timescales} the ${\rm H_2}$--cooling and dynamical 
time-scales, given by:
\begin{equation}
\tau_{\rm dyn} = \sqrt{\frac{3\pi}{16{\rm G}\rho}},  
\end{equation} 
\begin{equation} 
\tau_{\rm cool} = \frac{(3/2)n{\rm k_BT}}{|\Lambda_{\rm H_2}|},
\end{equation}  
as functions of radius in the five halo outputs. As in the
original simulations, we employ the ${\rm H_2}$--cooling rate, 
$\Lambda_{\rm H_2}$, provided by \citet{GP98}; here, G and 
${\rm k_B}$ are the gravitational and Boltzmann's constant, 
respectively. 

For all radii considered thus far, $0.1 < R < 10$ pc, Figure
\ref{fig:timescales} shows that the dynamical time exceeds the ${\rm
  H_2}$--cooling time in each of the cold haloes (as expected).  More
importantly, these time-scales are comparable -- i.e. within an order
of magnitude -- in both hot and cold haloes at $R \lsim 10$ pc (with
the exception of output 3, where the cooling time becomes much longer
outside the range $2 \lsim R \lsim 10$ pc). Because $\tau_{\rm
  cool}<\tau_ {\rm dyn}$ is required for ${\rm H_2}$--cooling to
impact the dynamics of these clouds, it is precisely in regions where
these time-scales are similar that inaccurate values of the
dissociation rate may qualitatively change the history of a simulated
cloud.  This will indeed be the case, as long as photodissociation and
shielding are both non-negligible.  For example, if the approximated
$f_{\rm sh}$ is too large and the ${\rm H_2}$--cooling rate therefore
erroneously low, the collapse of overdense clumps in the halo
(ultimately leading to formation of protostellar objects) may be
erroneously delayed. Conversely, if the adopted model overestimates
shielding, ${\rm H_2}$--cooling may allow for collapse sooner (on
smaller scales) than would be possible with the accurate dissociation
rate.
\begin{figure}
\includegraphics[clip=true,trim=.35in 4.in 3.5in 0.4in, height = 2.5in,width = 3.4in]{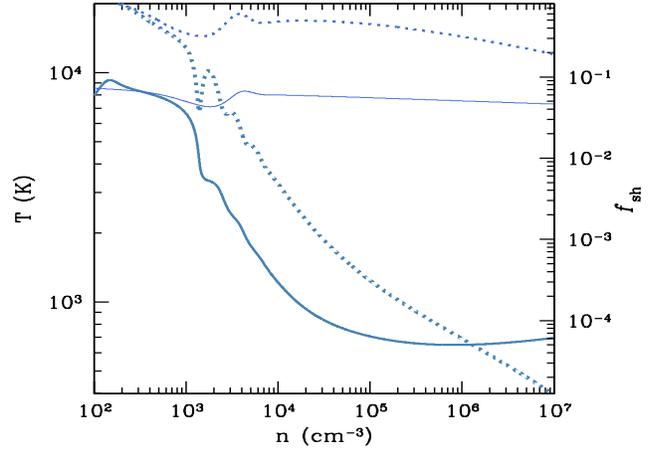}
\caption{Evolution of gas temperature (solid curves and labels on
  left) and $f_{\rm sh}$ (dashed curves and labels on right) of
  collapsing gas in a halo with $T_{\rm vir} \gsim 10^4$ K, are
  obtained from the spherical collapse model. Both are shown for runs
  with intensities, $J_{21}$, just above (thin curves) and below
  (thick curves) the critical threshold, $J_{\rm crit,21}$. The shield
  factor is obtained from the modified fitting formula,
  equation~(\ref{eq:newfit}), and $J_{\rm crit,21} = 1.4 \times
  10^3$. By contrast, if equation~(\ref{eq:DB37}) is used, $J_{\rm
    crit,21} = 4.3 \times 10^3$ and the bifurcation around threshold
  flux occurs at $f_{\rm sh} \approx 0.1$.  The incident spectrum is
  modeled as that of a blackbody with $T = 10^5$K and the temperature
  of the gas is initialized at the virial temperature of the halo. For
  more on the specifics of the spherical collapse model, the reader is
  referred to SBH10 and references therein.}
\label{fig:jcrit}
\end{figure}

In order to further address how errors in the approximate $f_{\rm sh}$
may impact the history of a simulated halo, we consider the ``critical
flux,'' studied by SBH10. It is well known that a sufficiently strong
LW flux can suppress ${\rm H_2}$--cooling entirely, keeping the gas in
these haloes close to its virial temperature, $T \approx 10^4$ K,
providing that it remains un-enriched in metals.  There is a sharp
bifurcation in the cooling behaviour around the threshold intensity,
$J_{\rm crit,21}$, below which the gas is able to reach $\sim$ a few
hundred Kelvin via ${\rm H_2}$--cooling (outputs 1, 2, and 5), while
haloes irradiated with a super-critical flux will remain ${\rm
  H_2}$--poor (outputs 3 and 4). From the full suite of simulations,
SBH10 found a critical intensity $J_{\rm crit,21}=10^{4-5}$, in the
usual units (here and throughout this discussion, the results are
quoted for an incident blackbody spectrum with $T_{\rm BB} = 10^5$ K,
relevant for direct ${\rm H_2}$ photodissociation). These results were
refined with the use of a one-zone spherical collapse model (for
details see SBH10, \citealt{OSH08}), and the threshold intensity was
then found to be $J_{\rm crit,21}=1.2 \times 10^4$.

To further investigate the implications of inaccurate self--shielding
models, we have repeated the one--zone calculations in SBH10, altering
only how $f_{\rm sh}$ is specified. First, we adopt the more accurate
expression given by DB96, equation~(\ref{eq:DB37}), rather than their
power--law fit, equation~(\ref{eq:DB36}), employed by SBH10. As in the
original study, the ${\rm H_2}$ column density is still specified by
the Jeans length (the only viable option for zero--dimensional models
without additionally computing gradients). The specific intensity,
$J_{\rm LW}$, is then varied iteratively with a Newton--Raphson scheme
and the critical threshold is found to be $J_{\rm crit,21} = 4.3
\times 10^3$. This reduction in the intensity required to suppress
cooling owes to the temperature dependence of
equation~(\ref{eq:DB37}), which more accurately models decreased
shielding due to line--broadening than the power--law. Next, we alter
the expression for $f_{\rm sh}$ as described in
\S~\ref{sec:FittingFormulae} (with $\alpha = 1.1$), and find $J_{\rm
  crit,21} = 1.4 \times 10^3$. Again, the more accurate fit at high
temperatures leads to significantly weaker shielding in the early
stages of collapse. The evolution of both temperature and the shield
factor as functions of density are shown in Figure \ref{fig:jcrit}.

Importantly, because we have employed the Jeans length to estimate
$N_{\rm H_2}$, the dissociation rate is still likely underestimated by
a factor of $\gsim 2$ at $f_{\rm sh} \lsim 10^{-3}$ (see the upper
left panel of Figure \ref{fig:Napprox}). As a result, the critical
intensity quoted here -- reduced by a factor of $\approx 3\times 2=6$
-- is still an {\it upper limit}; the full reduction is thus about an
order of magnitude.  Indeed, SBH10 showed that if one entirely ignores
self-shielding (i.e. the optically thin limit), then $J_{\rm crit,21}
= 4.4 \times 10^2$ -- apparently when self-shielding is treated more
accurately, it only modestly increases this value.

This reduction in $J_{\rm crit,21}$ illustrates how the
self--shielding approximations can dramatically impact the thermal
history of a simulated halo.  Furthermore, this may also have
interesting cosmogonical implications, suggesting that a larger
fraction of haloes than previously thought will see a supercritical
flux in the spatially fluctuating UV background (UVB). It has been
suggested (see SBH10 and references therein) that gas irradiated by a
supercritical flux may avoid fragmentation on small scales, provided
that it does not become metal--enriched \citep[see][for the relevant
metallicity threshold]{OSH08}, and this may provide a possible
mechanism by which primordial gas could collapse directly to form
massive black holes, $10^{4-6}~{\rm M_\odot}$.  However, with the
original high $J_{\rm crit,21}$ value, it has also been shown
\citep{Dijkstra+08} that only one in $\approx10^6$ haloes -- only those
with an unusually bright and close neighbour -- will see a
sufficiently high flux.  The direct collapse scenario requires that
the gas furthermore remain un--polluted by metals and efficiently shed
its angular momentum. As a result, it is likely that only a small
fraction of even these close halo pairs will form supermassive black
holes. The reduction of the $J_{\rm crit,21}$ value will significantly
increase the number of candidates for objects that avoid ${\rm
  H_2}$--cooling and fragmentation, and makes this scenario much more
viable.

\subsection{Internal and external modification of the 
radiation field by HI}
\label{sec:HI}

The results presented thus far may be altered by atomic H Lyman series
absorption either within the haloes, where this provides an additional
means of shielding ${\rm H_2}$, or in the IGM, where the spectrum of
the UVB is modified by the high HI optical depth prior to
reionization.

In the first case, suppression of the ${\rm H_2}$ dissociation rate
becomes significant if large HI column densities are present, $N_{\rm
HI} \gsim 10^{23}~{\rm cm^{-2}}$. To quantify the impact this has on
our results for $k_{\rm diss}$, we have repeated the
three--dimensional calculations with the first nine Lyman lines
contributing to the optical depth of each slab. Note that, while the
Ly$\alpha$ transition (10.2eV) lies outside of the LW wavelength range
($\gsim 11.2$eV), it is nevertheless included, because at large
$N_{\rm HI}$ its damping wings protrude into the LW band, and
contribute to LW shielding. 

In the SBH10 haloes, neutral column densities of $10^{21-22}~{\rm
  cm^{-2}}$ are typical at $R \sim 5-10$ pc, increasing to
$10^{23-24}~{\rm cm^{-2}}$ at radii smaller than a few pc. In these
densest regions, the dissociation rate is deceased by up to a factor
of $\sim 2.5$ in both cold and hot haloes. We find that the numerical
results for $f_{\rm sh} (N_{\rm H_2}, N_{\rm HI}, T)$ are well
approximated by an analytic expression of the form $f_{\rm sh,H_2}
\times f_{\rm sh,HI}$, where $f_{\rm sh,H_2}$ is the fit for
self--shielding in equation \ref{eq:newfit} (with $\alpha = 1.1$) and
$f_{\rm sh, HI}$ is the expression provided by \citet{WGH11} for HI
shielding:

\begin{equation}
f_{\rm sh, HI} = \frac{1}{\left(1 + x_{\rm HI}\right)^{{\rm \beta}}}
\times \exp\left({\rm-\gamma}\; x_{\rm HI}\right).
\label{eq:fHIfit}
\end{equation}

Adopting the same coefficients as in the original expression, 
$\beta = 1.6, ~\gamma = 0.15$ and $x_{\rm HI} = N_{\rm HI}/2.85 
\times 10^{23} {\rm cm^{-2}}$, we find that $f_{\rm sh,H_2} 
\times f_{\rm sh,HI}$ is accurate to within a factor of two in the 
relevant column density and temperature ranges: $10^{13} < N_{\rm H_2} 
< 10^{21}~{\rm cm^{-2}}, ~10^{22} < N_{\rm HI} < 10^{24} ~{\rm cm^{-2}}$, 
and $500 <T < 5 \times 10^3$ K. Note that this degree of accuracy is 
perhaps surprising, given that the fit $f_{\rm sh, HI}$ was originally
developed for HI shielding of ground state ${\rm H_2}$ populations;
for more details on this method for approximating the (non--linear) 
effect of combined shielding by ${\rm H_2}$ and HI, the reader is 
referred to \citet{WGH11}. 

The shape of the UVB spectrum is also modified by HI in the IGM, as
photons traveling over cosmological distances will be absorbed once
they redshift into resonance with a Lyman line. Subsequently, the
original photon will be replaced, in a radiative cascade, by several
lower-energy photons\footnote{In the case of Ly$\alpha$, the original
photon may only be lost through two--photon decays.  However, this
line is outside of the wavelength range of interest for ${\rm H_2}$
anyway.}.  This results in a characteristic sawtooth shape of the
spectrum (see \citealt{HRL97} and HAR00 for a detailed discussion). To
investigate how this modulation of the incident flux impacts our
results, we have recalculated $f_{\rm sh}$, assuming a single
(isothermal and static) slab of gas irradiated on one side, with the
input spectrum given by $J_\nu = J_{21} \times$ sawtooth modulation (z
= 15), adopted from HAR00 (from the model shown in their Figure 1). We
have found that the self--shielding behaviour is insensitive to this
modification -- $f_{\rm sh}$ is decreased by $\sim 10$ per cent over a wide
range of temperatures ($10^2-10^4$ K), column densities ($N_{\rm H_2}<
10^{24}~{\rm cm^{-2}}$), and irrespective of the assumed ${\rm H_2}$
level populations. While the SBH10 halo outputs are at redshifts
somewhat lower than $z = 15$, the magnitude of the sawtooth effect
decreases over time in the HAR00 model, due to the redshift--dependent
source formation rate; thus, the decrease in $f_{\rm sh}$ quoted above
is in fact an upper limit.\footnote{It is worth noting that the optically
thin rate is decreased by a factor of $\sim 10$ compared to that given
by the unmodified spectrum. However, we are primarily interested here
in the self--shielding behaviour, rather than in the changes to the
model--dependent magnitude of the intergalactic UVB.}

\subsection{Anisotropic LW flux} 
\label{sec:anisotropy}

In taking the average dissociation rate over all sight lines for the
3D calculations, we have thus far assumed isotropy of the incident LW
flux. This is likely a good approximation when the flux is close to
the global mean background, and is dominated by a large number of
distant sources.  However, if the radiation field is strong, i.e. the
halo has one (or a few) bright neighbours, then the flux will be
anisotropic, with $f_{\rm sh}$ for individual sight lines pointing
toward the nearby source most important. Indeed, the latter case is
particularly relevant in the context of the critical flux discussed
above, as supercritical intensities, $J_{\rm 21} \gsim 10^3$, are
likely to be well above the cosmic mean \citep{Dijkstra+08}.

\begin{figure}
 \includegraphics[clip=true,trim=0.2in 4in 3.65in 0.4in,                    
height = 2.5in,width = 3.4in]{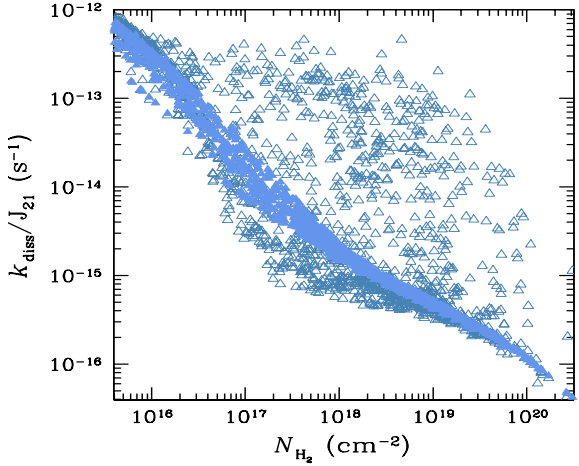}                                            
\caption{Comparison of the numerically calculated photodissociation rate 
  coefficient along individual sight lines emanating from $~\sim$ 100 points 
  at $1 < R < 10$ pc in output 5 (open triangles) with those obtained
  when the gas is assumed to be both static and isothermal (closed
  triangles).}
\label{fig:anisotropic}
\end{figure}

Let us consider the ramifications of anisotropic incident flux
for self--shielding models in the context of gas velocity and
temperature gradients. As shown in Figure~\ref{fig:sightlines}, the
shield factor for a given sightline may be altered dramatically by
variations in the gas velocity or temperature in that direction. This
point is further illustrated in Figure \ref{fig:anisotropic}, in which
$f_{\rm sh}$ for individual sight lines emanating from 100 points are
shown in comparison with those when both velocity and temperature
gradients are artificially switched off. (See
\S~\ref{sec:vgradients}--\ref{sec:Tgradients} for details of how the
observed discrepancies arise.) Here, we see that the dissociation rate
fluctuates by $\approx$two orders of magnitude in different
directions.  The rates will also be incorrect by up to several orders
of magnitude if one makes the standard assumption of static and
isothermal shielding gas.  (On the other hand, the relative accuracy
of each $N_{\rm approx}$ along individual sight lines would not be
particularly informative here, largely because simulations in which
the radiation field is both spatially and angular resolved typically
do not rely on the type of local estimates for $N_{\rm H_2}$ described
above.)

\subsection{Uncertainty in the rovibrational distribution}
\label{sec:LevelPopulations}
\begin{figure*}
\includegraphics[clip=true,trim=0in 3.8in 0in 0.4in,
height = 2in,width = 4.8in]{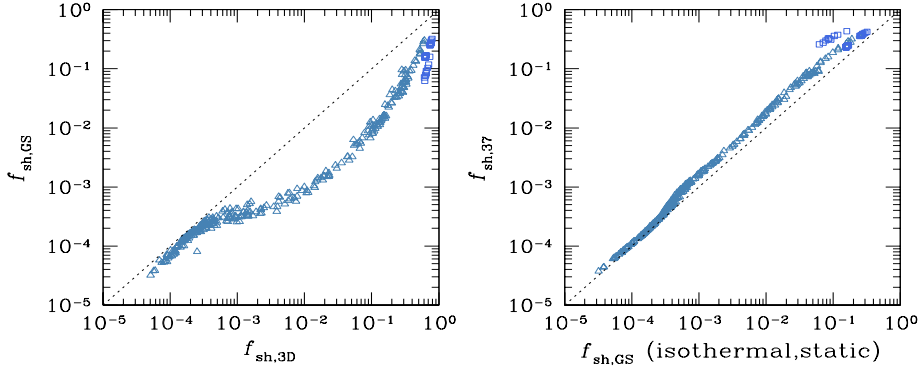}
\caption{{\it Left panel:} comparison of full three--dimensional
  calculations with LTE populations, $f_{\rm sh,3D}$, to those in
  which only the ortho and para ground states of the molecules are
  populated, with a fixed ortho:para ratio of 3:1. {\it Right panel:}
  results of the ground--state calculation are compared to those from
  the DB96 fitting formula, equation~(\ref{eq:DB37}). In the latter
  case, the mean ($f_{\rm sh,37}$) over all 16 sight lines is shown,
  with velocity and temperature gradients artificially switched off,
  in order to isolate the accuracy of the fitting formula.}
\label{fig:groundstate}
\end{figure*}

In the most general case, one would have to consider the
non-equilibrium ${\rm H_2}$ population levels produced during
gas--phase ${\rm H_2}$ formation, as well as the time--dependent
cascades among levels.  This, in general, would produce a population
level distribution that is different than assumed in DB96. Absent such
a fully time--dependent calculation of the rovibrational level
populations, we have thus far made the simplifying assumption that
rotational states within the ground vibrational state are populated
according to a Boltzmann distribution and that the abundance of ${\rm
  H_2}$ in higher vibrational states are negligible. However, because
the level populations play an important role, particularly in the
temperature dependence of the dissociation rate (see
\S~\ref{sec:FittingFormulae}-\S~\ref{sec:Tgradients}), a more detailed
examination of this assumption is in order.

It has already been noted that thermal populations are not established
in higher vibrational levels until particle densities $n_{\rm crit}
\sim 10^{6-7}~{\rm cm^{-3}}$ are reached (see, e.g. Table 1 in LPF99),
so there is little uncertainty that populations in $v > 0$ are
negligible for the SBH10 haloes, at least at the most relevant stages
of collapse that we considered here. However, higher rotational levels
within $v = 0$ also thermalize at higher critical densities, and
because the collisional cross-sections depend strongly on temperature,
so also do the values of $n_{\rm crit}$. As an example, consider the
S(2) transition (for $v =0$), for which the critical density at
500, 1000, and 2000 K, respectively, is $n_{\rm crit} = 1.2 \times
10^4,~1.7\times 10^3,~2.5 \times 10^2~ {\rm cm^{-3}}$ (LPF99). Given
that the most rarefied regions of the haloes $n < 10^2~{\rm cm^{-3}}$
are typically at temperatures of $\sim$ several thousand Kelvin (see
Figure \ref{fig:profiles}), most likely the rotational populations
within $v=0$ will indeed tend to a Boltzmann
distribution. None the less, we have repeated our three--dimensional
calculations with populations in $J=0,1$ only, and with a fixed
ortho:para ratio of 3:1. The results are compared to the original
$f_{\rm sh,3D}$ in the left panel of Figure \ref{fig:groundstate}.
Because restricting the molecules to occupy only $J = 0,1$ is, in
effect, similar to setting the temperature very low ($T \lsim 100$ K)
in the LTE model, it is not surprising that the shield factor is
smaller in this case. The clear dip in the results above $f_{\rm
  sh,3D} \gsim 5 \times 10^{-4}$ arises simply because the optical
depth decreases toward hotter regions of these haloes. At the smallest
values of $f_{\rm sh}$, the local temperature is only a $\sim$ few
hundred Kelvin, and as a result, the ground state results do not
differ dramatically from the original (LTE) calculations.  In the
right panel of Figure \ref{fig:groundstate}, the ground state results
are also compared to those from equation~(\ref{eq:DB37}), confirming
that the DB96 expression is indeed a better fit to the numerical
results in a rotationally--cold gas.

\subsection{Fluorescent excitation of ${\bf H_2}$}

All calculations to this point have implicitly assumed that each LW
photon is permanently removed from the radiation field upon absorption
by ${\rm H_2}$.  However, on average only $10-15$ per cent of absorption
(``pumping'') events result in dissociation, while the remainder are
followed by transitions to another bound vibrational state. In a
``resonant scattering,'' a single decay returns the molecule directly
to the initial rovibrational state ($v,J = v'',J''$) and the original
LW photon is re--emitted. More frequently, UV--pumping is followed by
a cascade through multiple levels, resulting in both infrared
fluorescence and emission of a LW photon (with different-energy) in
the electronic transition. It is then possible for this (or the
resonantly scattered) photon to be reabsorbed, a process which has not
been accounted for thus far. If the optical depth to the re--emitted
photon is non--negligible this requires a non--trivial modification in
our calculations of $k_{\rm diss}$. However, recall that only ${\rm
  H_2}$ transitions originating in the $v=0$ are included; as a
result, to first order, we need only consider photons emitted in
decays directly back to the ground vibrational state. The fraction of
downward transitions to each bound vibrational level are quantified by
\citet{Shull78}; they find that $\sim 15$ per cent of all absorption events
originating in $v=0$ result in decays directly back to $v''=0$ (see
their Table 1). Therefore, accounting for the optical depth to
re--emitted photons represents a small correction to our results.

There is an additional complication, however, if the gas is irradiated
by a very strong UV flux, in which the cascade to $v''=0,J''$ may be
interrupted by absorption of LW photons from $v'' > 0$. However, only
in very strong radiation fields, $J_{21} > 10^{5-8}$, are molecules
more likely to be ``re-pumped'' in this way, rather than undergoing
radiative transitions to lower rovibrational states \citep{Shull78}.
Therefore, in the present context of an intergalactic UVB, neglecting
this multiple pumping mechanism is also justified.

\section{Summary and Conclusions}
\label{sec:Conclusions}

We have shown that the results of existing approximations for
self--shielding in three--dimensional simulations often introduce
large inaccuracies in the optically thick H$_2$ photodissociation rate. In particular, the
approximate results typically underestimate the numerically calculated
rate by more than an order of magnitude in low density regions, $n <
10^4~ {\rm cm^{-3}}$, or where the true shield factor is $10^{-3}
\lsim f_{\rm sh,3D} \lsim 1$. At higher densities, we find that the
approximate methods are much more accurate, with typical errors of
$\sim 25$ per cent. There are a number of factors contributing to the
discrepancies between the approximate and numerical results,
which are summarized below:\\

\begin{enumerate}
\item The largest source of error in the approximate methods is the
  analytic fit for $f_{\rm sh}$ provided by DB96. While this oft--used
  expression is reasonably accurate at low temperatures, we find that
  it overestimates shielding by a large factor at temperatures above a
  few hundred Kelvin (for a gas in LTE). The resulting deviations from
  our numerical results are most apparent at $n < 10^4~{\rm cm^{-3}}$,
  as these low--density regions typically have not cooled below $\sim
  500- 10^3$ K in the SBH10 simulations. However, we have found that a
  very simple modification to the DB96 expression (see equation
  \ref{eq:newfit}) improves the agreement with our numerical results 
  to within $\sim 15$ per cent.\\

\item Nearly all existing approaches to approximate $f_{\rm sh}$ are
  based on a static slab model for the shielding gas, which neglects
  the diminished optical depth due to frequency shifts of the ${\rm
    H_2}$ resonances in the presence of velocity gradients. We find that
  these frequency shifts do not greatly alter the dissociation rate in
  the SBH10 haloes, largely because the gas motions in these
  simulations are dominated by a coherent flow toward the
  dense core, rather than (supersonic) turbulence; therefore,
  typically only a small number of sightlines from a given point see
  large changes in the bulk velocity. The resulting increase in the
  spherically averaged rate is only significant ($\sim 65$ per cent)
  in the range $4 \times 10^{-4} \lsim f_{\rm sh} \lsim 2 \times
  10^{-2}$,
  and elsewhere is generally negligible. \\

\item We find that gas temperature gradients can alter the
  self--shielding behavior dramatically via changes in both the
  thermal line widths and in the rovibrational level populations. In
  this case, the dissociation rate may be either increased or
  decreased, depending on the sign of the temperature gradient along
  each sightline. However, the effect again is smaller for the
  spherically averaged rate, and tends to make shielding stronger in
  the haloes we analyze, largely counteracting any changes due to
  frequency shifts.  Taken together, temperature and velocity
  gradients therefore typically only introduce a factor of $\lsim 2$
  scatter in spherically averaged rate. \\
 
\item Finally, we have evaluated several approaches to estimate the
  shielding column density in simulations (and one--zone models). The
  most common of these is based on the assumption that the
  characteristic length scale for shielding is of order the local
  Jeans length. In agreement with SBH10, we find that this method is
  very accurate at densities $n \gsim 10^{4-5} ~{\rm cm^{-3}}$, but
  underestimates the optically thick rate by $\gsim$ an order of
  magnitude at lower densities.  However, we show that two less
  commonly used methods provide reasonably accurate estimates of the
  shielding column at all densities we consider, and are
  computationally inexpensive, relying again only on local properties
  of the gas. These are based on the Sobolev length and a variation of
  the Sobolev length based on the density -- rather than velocity --
  gradient; both yield more accurate results for $f_{\rm sh}$ than the
  Jeans length method in low--density regions, with essentially
  unbiased scatter around the true value.  We also show that a
  ``six--ray'' method, based on integrating the column density in only
  six directions is extremely accurate, deviating from the ``exact
  results'' only due to the effects of temperature and velocity
  gradients. However, this non--local method comes with a larger 
  computational expense.

\end{enumerate}

In addition to the factors enumerated above, HI shielding of ${\rm
  H_2}$ also causes the ``true'' dissociation rate to deviate from the
results of approximate treatments, which most often neglect this
additional shielding.  We find that this effect can be well modeled by
a simple analytic prescription provided by \citet{WGH11}.  The simple
fitting formulae we provide can be trivially incorporated into future
three--dimensional simulations, to improve the speed and accuracy of
calculations of the ${\rm H_2}$--photodissociation rate.

Finally, using the same (one--zone) spherical--collapse model employed
by SBH10, we show that the critical LW flux $J_{\rm crit}$ required to
keep $T_{\rm vir} \gsim 10^4$K haloes ${\rm H_2}$--poor is reduced by
about an order of magnitude as a result of improved accuracy in the
shielding factor. This serves to illustrate an important conclusion of
our study: the cooling properties -- and thereby the dynamical history
-- of (metal--free) gas in simulated haloes depend sensitively on the
adopted self--shielding model.  In particular, the reduction in
$J_{\rm crit}$ implies that ${\rm H_2}$--cooling is suppressed in many
more of these haloes, thus increasing the potential sites for direct
formation of SMBHs in the early universe.

\section{Acknowledgments} 
We thank Cien Shang for helpful discussions, and for sharing the
simulation data. We are also grateful to the developers of the
analysis toolkit {\sc yt} (http://yt.enzotools.org), which facilitated
analysis and visualization of the {\sc enzo} data.  We acknowledge 
support from NSF grants AST-05-07161, AST-05-47823, AST-09-08390, 
and AST-10-08134, as well as computational resources from NASA, 
NSF Teragrid, and Columbia University's Hotfoot cluster.

\bibliography{H2}

\end{document}